# Limits of reductionism and the measurement problem

Arkady Bolotin[a]

*Ben-Gurion University of the Negev, P.O.B. 653, Beersheba 85105, Israel*

ABSTRACT

In the paper, we tackle the following questions: Could the difficulty in solving the Schrödinger equation for an arbitrarily large system be a reflection of some nature's intrinsic property? And if so, could this difficulty be a resolution to the measurement problem?

RÉSUMÉ

Dans le papier, nous abordons les questions suivantes : La difficulté en résolvant l'équation de Schrödinger pour un système arbitrairement grand a-t-elle pu être une réflexion de la propriété intrinsèque d'une certaine nature? Et si oui, cette difficulté a-t-elle pu être une résolution au problème de mesure?

---

[a] arkadyv@bgu.ac.il

"The general theory of quantum mechanics is now complete... The underlying physical laws necessary for the mathematical theory of a large part of physics and the whole of chemistry are thus completely known and the difficulty is only that the exact application of these laws leads to equations much too complicated to be soluble. It therefore becomes desirable that approximate practical methods of applying quantum mechanics should be developed, which can lead to an explanation of the main features of complex atomic systems without too much computation." – Paul Dirac, *Quantum Mechanics of Many-Electron Systems*[1]

## 1. Introduction

In accordance with inter-theoretic reductionism, in principle, any physical system can be described by the many-body Schrödinger equation as long as the constituent microscopic particles are not moving "too" fast; that is, if they are not moving near the speed of light. This covers a wide range of problems, so if we could manage to solve the Schrödinger equation, not only we would be able to predict the behavior of molecular, atomic, and subatomic systems, but also macroscopic ones, and possibly even the whole universe.

Still, despite the fact that more than two generations of researchers were left to work out how to achieve this ambitious goal, the only exact solutions to the Schrödinger equation found so far are for free-particle motion, the particle in a box, the hydrogen atom, hydrogen-like ions, the hydrogen molecular ion, the rigid rotator, the harmonic oscillator, Morse and modified Morse oscillators, and a few other systems[2,3]. For more complicated systems, however, approximation techniques have to be used (such as the variational method or perturbation theory), which sometimes give poor results compared with experimental ones, and practical calculations with them are usually very difficult, even with the use of powerful computers[4]. The difficulty is that in a system made of $N$ interacting particles (where $N$ can be anywhere from three to infinity), the repeated interactions between particles create quantum correlations. As a consequence, the dimension of the Hilbert space describing the system scales exponentially in $N$. This makes a direct numerical calculation of the Schrödinger's equation intractable: Every time an extra particle is added to the system, the computational resources would have to be doubled[5].

On the other hand, computational power is increasing constantly. As we solve larger and more complex problems with greater computational power and cleverer algorithms day after day, we may assume that eventually we will be able to solve the many-body Schrödinger equation for a system of arbitrary complexity (at least, no law or known impossibility result in computational complexity theory can exclude such an assumption of infinite computational power).

So, those, who criticize reductionism for being unable to solve problems in chemistry, biology, economics or psychology, might really have a problem only with its manifestation as a current working principle and not with reductionism itself, which is in theory capable to account for any higher-order phenomenon. The present-day reductionism has failed to reduce every explanation in every field of science to quantum mechanics probably because of temporary difficulties with solving the Schrödinger equation. Therefore, a hierarchical or layered view of the whole of nature, with the layers arranged in terms of increasing complexity with each requiring its own special science[6], might turn out to be just a usable concept pending future efficient solutions to the many-body Schrödinger equation.

Critiques of intertheoretic reductionism from the perspective of emergentism or holism emphasize a high amount of complexity in a system as the main argument setting a limit to reductionism[7]. Systems in chemistry, biology, psychology, or sociology, they argue, are commonly so complex that it will not ever be possible to describe all their details[8]. Those systems are so complex that their behavior is, or appears, "new" or "emergent": It cannot be predicted from the properties of the elements alone[9]. Then again, complexity as an argument against reductionism might be valid only in the interim, i.e., at the current level of computation[10]. If the number of calculations required to reach a solution for any given mathematical problem could be written as some polynomial function of the problem's input size, then with a substantial computational power one would be able to describe any complex system in full detail and completely predict its behavior from the Schrödinger equation alone.

It may seem that intertheoretic reductionism in company with infinite computational power would know no limit. And yet the infamous measurement problem in quantum mechanics might prove a serious hindrance even for it.

## 2. Measurement problem
## and the assumption of infinite computational power

Briefly, the quantum measurement (or macro-objectification) problem runs as follows. The wavefunction in quantum mechanics evolves according to the Schrödinger equation into a sum (superposition) of different states, but the application of this principle to macroscopic systems appears to lead immediately to a disagreement with our everyday experience[11].

The assumption of infinite computational power merely makes this problem unambiguous. Indeed, if we assume that with a computer we may solve any given mathematical prob-lem in a reasonable amount of time, we won't be able to claim any longer that both the Schrödinger equation and its solution for an arbitrary macroscopic system can be presented only symbolically, i.e., in purely abstract terms, lacking any considerable content, even in the face of the internal complexity of the macroscopic system and its external environment. In that case, to paraphrase Schrödinger[12], nothing would stop ''indeterminacy originally restricted to the atomic domain" becoming transformed into "a macroscopic indeterminacy": The linearity of the Schrödinger equation combined with infinite computational power would inevitably bring forth an actual and hence testable solution allowing the macroscopic system to be in a superposition of macroscopically distinguishable states.

Therefore, by assuming that we can compute the Schrödinger equation for any system in a reasonable time, we make the quantum description of a truly macroscopic object feasible and consequently obtain a situation where the theoretical inference clashes with the evidence: The calculated macroscopic quantum superpositions are at odds with the observed behavior of the macro-object.

What are the ways out of this dilemma? In general, we may think of two principal routes leading from it.

The first route is *to insist on unrestricted, limitless reductionism* and hence remaining within the framework of the orthodox or modified quantum mechanics find out why macroscopic quantum superpositions are not observable.

The second route is *to admit the limitation of reductionism* and thus standing beyond the formalism of quantum mechanics find out why the quantum description is not applicable to macroscopic objects.

The first route, meticulously studied and discussed in many books and countless papers, contains suggestions of changing the interpretation of quantum mechanics whilst leaving the Schrödinger equation unchanged (which includes ideas from the many-worlds interpretation and the decoherence program to the one that physical reality is the product of discrete subphysical information processing equivalent to the actions of a probabilistic Turing machine[13]), in common with alternative suggestions of injecting new physics by modifying the Schrödinger equation (with, for example, an additional small, nonlinear contribution[14]).

As to the second route, it is virtually unexploited other than some thoughts expressed by a few researchers now and then, which argued in favor of such an approach but did not elaborate it at length.

Thus, for instance, Bohr[15] has argued that classical physics does not emerge from quantum physics as an approximation of the latter when $\hbar$ tends to zero.[b] His position is that classical concepts are autonomous from quantum theory and cannot be derived from it. As stated by Bohr's correspondence principle (in its strong form), it is inappropriate to treat macroscopic measuring instruments in purely quantum mechanical notions.

In a similar vein, Ruža[18] argued that "despite our theoretical views and assumptions, it seems that the perceived outside world is really split into two domains — the quantum one and the classical one. Every domain has its own rights to exist, neither has an ontological priority over another, i.e., there are no substantial, unconditional arguments for the primacy of quantum physics over classical".

But what is the reason for such a split between domains that poses a limit to reductionism?

---

[b] The problem of the emergence of classical mechanics from quantum mechanics is still open. In spite of many results on the $\hbar \to 0$ asymptotics, it is not yet clear how to explain within standard quantum mechanics the classical motion of macroscopic bodies[16]. The limit $\hbar \to 0$ transforms quantum mechanics into a classical probabilistic theory[17].

As an explanation or justification for this split, we may adopt an idea just opposite to what we took on earlier: We may assume that computational power is actually limited, viz., that there exists a limit of what could be solved efficiently with a computer no matter how fast and powerful it might be.

### 3. Assumption of limited computational power

Namely, we will assume that solving the many-body Schrödinger equation in its most general form is such a mathematical problem that has no *tractable computational solution*. By "tractable computational solution", we mean a solution that even in the worst case could be reached in the number of calculations $T(n)$ not greater than a polynomial expression of the problem's input size $n$ (which corresponds to the system's number of degrees of freedom), i.e., $T(n) \in O(n^k)$ for some constant $k > 1$. Let us briefly examine possible consequences of this assumption.

**1.** According to the assumption, no fast solution to the Schrödinger equation for an arbitrary system of $N$ interacting microscopic particles would be ever known; hence, a solution to the many-body Schrödinger equation would always require, at most, a superpolynomial number of calculations $T(n) > n^k$ for all $k$. That is, the time required to solve the many-body Schrödinger equation would increase very quickly as the number of constituent particles $N$ would grow. As a result, the time needed to reach a solution for a macroscopic system (i.e., the system which contains a number of constituent particles proportional to the Avogadro con-stant $N \propto 10^{23}$) could easily spread into the billions or trillions of years, using any amount of computing power available.[c]

This would make the quantum mechanical formalism inapplicable to a truly macroscopic object: In any reasonable amount of time, this formalism would not be able to give to such an object a deterministic description that could be supported or falsified by the data of actual experience.

**2.** Therewithal, the given assumption does not exclude the possibility that for some particular systems comprised of quite large numbers of particles there may be solutions to the Schrödinger equation reachable in realistic time. In addition, this assumption only relates to worst-case numbers of calculations; therefore, for certain systems the average number of calculations taken to reach the solution on a mesoscopic or even a macroscopic scale may possibly be reasonable.

This may explain why neither mass, nor length, nor any other physical property of a physical object can serve as a stringent criterion for the boundary between quantum and classical descriptions.

**3.** Further, the given assumption only considers numbers of calculations necessary to reach deterministic solutions to the Schrödinger equation. This suggests that a solution, which allows two-sided random errors in the solution's parameters, might be reachable in a reasonable number of calculations (even though the deterministic solution is not). In other words, it might be possible to reach a solution to the Schrödinger equation for a given system quickly – and therefore to make the quantum-mechanical description of this system feasible – at the cost of statistical uncertainty.[d]

---

[c] One may reply that quantum computers could simulate quantum physical processes exponentially faster than classical computers and accordingly they might solve the quantum many-body problem efficiently. However, no proof exists yet for the general superiority of quantum computers over their classical counterparts[19]. Specifically, quantum computers are neither known nor believed to be able to solve **NP**-complete problems efficiently[20].

This implies that large systems (such as observers and their measuring apparatuses) might be described in quantitative terms only by non-deterministic solutions to the Schrödinger equation. It may resolve the conflict between the deterministic dynamics of quantum mechanics and the postulate that during measurement a non-deterministic collapse of the wave packet occurred (the postulate of collapse).

## 4. Postulate of collapse
## and the assumption of limited computational power

Within the frame of the assumption of limited computational power (i.e., the intractability of the many-body Schrödinger equation), in most cases that involve solely microscopic systems, the deterministic solution to the Schrödinger equation might be reachable in realistic time (due to the small size of the problem). Consequently, quantum description of these systems could be valid: They would be completely described by the corresponding wave function that evolves gently, in a perfectly predictable and continuous way (that is, knowing the wave function at one moment, the Schrödinger equation determines – through a reasonable number of calculations – the wave function at any later time). However, as soon as a measurement is performed, a macroscopic system is involved, for which the Schrödinger equation has no tractable deterministic solution. Therefore, for a measuring array (composed of a measured microscopic system interacting with a macroscopic measuring device) a purely quantum description consisting of the Hilbert space tensor product of the state spaces associated with the microscopic system, the measuring device, and the totality of existence would be unfeasible (and accordingly inexpressible in quantitative terms). Instead, a quantum-mechanical description of the measuring process might be only statistical, i.e., one that is inexact (coarse-grained), providing with errors. In the aggregate, these errors might prevent different elements in the quantum superposition of the measured microscopic system's wavefunction from interfering with each other, giving the appearance of wavefunction collapse.

---

[d] Speaking in general terms, letting uncertainty in the system's solution can in effect make some degrees of freedom of the system irrelevant to the solution and hence possible to be skipped, which in turn can significantly lessen the amount of calculations needed to find the answer to such a simplified problem.

The next simple example illustrates this statement (the example follows the explanation of the quantum decoherence given in the paper by Dass[21] but assumes the intractability of the many-body Schrödinger equation). Consider a spin-$1/2$ particle (the test-particle) prepared in the state

$$|\psi_{x+}\rangle = \frac{1}{\sqrt{2}}(|\uparrow\rangle + |\downarrow\rangle) \quad (1)$$

as it passes through a reversible Stern-Gerlach apparatus which first splits the particle's trajectory into upward and downward branches that are correlated with the spin component of the particle and next recombines the branches before the particle leaves the experiment. A macroscopic detector is located near the upward branch and acts as a measurement device. Let us find what observers would perceive measuring the particle's spin after the particle has left the experiment.

To provide the description of the particle flying from the apparatus after having interacted briefly with the detector M and hence the entire macroscopic environment ENV, a solu-tion $|\psi(t)\rangle$ to the Schrödinger equation for this particle might only be inexact (lest the exact solution involving the entangled states $|\phi^{(\uparrow)}(t)\rangle_{\text{M+ENV}}$ and $|\phi^{(\downarrow)}(t)\rangle_{\text{M+ENV}}$,

$$|\psi(t)\rangle = \frac{1}{\sqrt{2}}\left(|\uparrow\rangle \otimes |\phi^{(\uparrow)}(t)\rangle_{\text{M+ENV}} + |\downarrow\rangle \otimes |\phi^{(\downarrow)}(t)\rangle_{\text{M+ENV}}\right) \quad , \qquad (2)$$

should be infeasible as possessing a great number of degrees of freedom). With that purpose, we accept two-sided random errors $\Delta u$ and $\Delta d$ in the definite energies $u$ and $d$ of the interaction between the particle and the detector associated with the upward and downward branches of the trajectory that the particle can take. If the resultant loss of information about the interaction energies $u$ and $d$ would be maximal (that is, if the absolute values of the errors $\Delta u$ and $\Delta d$ would be of the order of the interaction energies $u$ and $d$), then the interaction Hamiltonian $H_{\text{int}}$ (diagonal in the position-spin-space basis as the interaction is due to the Coulomb force) would take the stochastic form containing only the degrees of freedom of the test-particle[e]: $H_{\text{int}} \simeq (\bar{u} + \Delta u)|\uparrow\rangle\langle\uparrow| + (\bar{d} + \Delta d)|\downarrow\rangle\langle\downarrow|$; and so the state vector of the particle exiting

---

[e] The exact interaction Hamiltonian $H_{\text{int}}$ would enclose operators $A_u$ and $A_d$ acting on the Hilbert space of the quantum states of the detector and the environment $H_{\text{int}} = |\uparrow\rangle\langle\uparrow| \otimes A_u + |\downarrow\rangle\langle\downarrow| \otimes A_d$. For simplicity, if one treats the particle and the detector as a closed system, the $H_{\text{int}}$ would take the form

$$H_{\text{int}} = (|\uparrow\rangle\langle\uparrow|) \otimes \left(\sum_k u_k |\epsilon_k\rangle\langle\epsilon_k|\right) + (|\downarrow\rangle\langle\downarrow|) \otimes \left(\sum_k d_k |\epsilon_k\rangle\langle\epsilon_k|\right) \quad ,$$

where $|\epsilon_k\rangle$ represent the detector's orthonormal quantum states. The given form of the $H_{\text{int}}$ indicates that each configuration of the detector's microscopic constituent particles (like their positions) is characterized by the certain energy of the interaction – $u_k$ or $d_k$ – for each path that the test-particle can take. However, because of too numerous microscopic degrees of freedom of the detector, the exact solutions $|\epsilon_k\rangle$ to the detector's Schrödinger equation are infeasible to compute. Therefore, for the test-particle having interacted with the detector and then flying away we must be content with an approximate (coarse-grained) description that ignores the detector's microscopic degrees of freedom. To ob-from the apparatus $|\psi(t)\rangle$ would be given by the effect $U(t)|\psi_{x+}\rangle = |\psi(t)\rangle$ of the stochastic unitary time-evolution operator $U(t) = \exp(-itH/\hbar)$, where $H \simeq H_{\text{int}}$:

$$|\psi(t)\rangle \in \left\{ |\Delta u| \sim \bar{u}, |\Delta d| \sim \bar{d}: \ \frac{1}{\sqrt{2}} \left( |\uparrow\rangle e^{-i\bar{u}t/\hbar} e^{-i\Delta u t/\hbar} + |\downarrow\rangle e^{-i\bar{d}t/\hbar} e^{-i\Delta d t/\hbar} \right) \right\} \ . \quad (3)$$

To aggregate these non-deterministic solutions (3) we estimate the expected value of the probability[f] $P(|\psi_{x+}\rangle \to |\psi(t)\rangle) = |\langle \psi_{x+}|\psi(t)\rangle|^2$ of the transition from the initially prepared state $|\psi_{x+}\rangle$ to the possible final state $|\psi(t)\rangle$:

$$\overline{|\langle \psi_{x+}|\psi(t)\rangle|^2} = \frac{1}{2} + \frac{1}{2} \cos\left( \frac{\bar{u} - \bar{d}}{\hbar} t \right) \overline{\cos\left( \frac{\Delta u - \Delta d}{\hbar} t \right)} \ . \quad (4)$$

Denoting the random value $(\Delta u - \Delta d)t/\hbar$ by $\zeta$ and substituting $\zeta_{max} = -\zeta_{min} \sim (\bar{u} + \bar{d})t/\hbar$, we get the expression

$$\overline{\cos(\zeta)} = \frac{\hbar}{2(\bar{u} + \bar{d})t} \int_{-\hbar^{-1}(\bar{u}+\bar{d})t}^{\hbar^{-1}(\bar{u}+\bar{d})t} \cos(\zeta)\, d\zeta \underset{\bar{u},\bar{d} \ggg 0}{\simeq} 0 \ , \quad (5)$$

which is about zero as the estimated interaction energies $\bar{u}$ and $\bar{d}$ (approximately proportional to the number of electrons in the detector) tend toward high values in the macroscopic limit. This yields the aggregate result $P(S_x = +\hbar/2) \simeq 1/2$, meaning that after the detector "observes" the particle, the subsequent measurement of the particle's spin along the $x$-axis would find the eigenvalue $+\hbar/2$ almost half the time. In contrast, in the case of no interaction (i.e.,

---

tain such a description we should consent to the maximum uncertainty about the interaction energies $u_k$ and $d_k$ such that $u_k \in \{|\Delta u| \sim \bar{u}: \ \bar{u} \pm |\Delta u|\}$, $d_k \in \{|\Delta d| \sim \bar{d}: \ \bar{d} \pm |\Delta d|\}$: This would make the specifics about the individual configurations of the detector's microscopic particles immaterial to the inexact solutions $|\psi(t)\rangle$, that is $H_{\text{int}} \simeq |\uparrow\rangle\langle\uparrow|(\bar{u} \pm |\Delta u|)\hat{1} + |\downarrow\rangle\langle\downarrow|(\bar{d} \pm |\Delta d|)\hat{1}$ (where $\hat{1}$ denotes the identity operator).

[f] It should be noted that we could not directly aggregate the solutions (3) because – as it has shown in the paper by Sexton and Jones[22] – the arithmetic mean cannot be extended to complex numbers.

zero interaction time $t$ or zero interaction energy), we would get $P(S_x = +\hbar/2) = 1$, which means that the particle is not affected and therefore retains its spin. The analogous aggregate result $P(S_z = +\hbar/2) = 1$ would be ascertained if we let the particle in the eigenstate $|\uparrow\rangle$ pass through the apparatus: The state $|\psi(t)\rangle$ of this particle defined by the propagator $U(t)|\uparrow\rangle$ is an element of the following set

$$|\psi(t)\rangle \in \left\{ |\Delta u| \sim \bar{u}: \ |\uparrow\rangle e^{-i\bar{u}t/\hbar} e^{-i\Delta ut/\hbar} \right\} , \qquad (6)$$

subsequently the set of the transition probabilities $P(|\uparrow\rangle \to |\psi(t)\rangle) = |\langle\uparrow|\psi(t)\rangle|^2$ would be

$$|\langle\uparrow|\psi(t)\rangle|^2 \in \left\{ |\Delta u| \sim \bar{u}: \ \left| e^{-i\bar{u}t/\hbar} e^{-i\Delta ut/\hbar} \right|^2 = 1 \right\} . \qquad (7)$$

Thus, allowing only feasibly computable (and therefore testable) solutions to the Schrödinger equation might reproduce approximately classical (i.e., probabilistically additive) results of the measurement and preclude details of the measuring device from appearing in the description of the measured system.[g]

## 5. Measurement problem and P versus NP problem

Intuitively it is clear that once the solution $\psi(\boldsymbol{r}_1 \ldots \boldsymbol{r}_N, s_{z1} \ldots s_{zN}, t)$ to the Schrödinger equation for a system of $N$ interacting microscopic particles (each with spin) is known, to verify it would require a number of steps (i.e., observations or experiments), in any case, not greater than a polynomial of $N$.

---

[g] This would correspond to Streater's remarks[23]: "The idea that the full details of the observer should be included in the Hilbert space is in violation of the scientific ethos… The theory needs a cut, between the observer and the system, and the details of the apparatus should not appear in the theory of the system".

This entails that finding the function $\psi$ that solves the Schrödinger equation is the mathematical problem in **FNP** complexity class (which is the function-problem extension of the decision-problem class **NP**). Roughly speaking, **FNP** is the class of functions that can be verified efficiently (i.e., quickly – in a polynomial number of steps), whereas **FP** is comprised of the class of functions that can be computed efficiently on a classical computer without randomization[24]. As the assumption of limited computational power states, the wave function $\psi$ is outside of the **FP** class.

It is also intuitively clear that if the solution for a particular natural-world problem in any of the special sciences – chemistry, biology, neuroscience or others – can be expressed in the form of a function of the problem's inputs $q_1 \ldots q_M$ (i.e., input or independent variables), then the given solution $\mathcal{F}(q_1 \ldots q_M, t)$ would be easy to verify (i.e., it can be verified in a number of steps polynomial of $M$). Accordingly, this particular problem would be in the **FNP** complexity class.

Because of the Schrödinger equation's fundamentality (this equation is supposed to govern all natural processes), each natural-world problem in the **FNP** class can be reducible to the problem of solving the Schrödinger equation for the corresponding underlying quantum model by replacing the natural-world problem's inputs $q_1 \ldots q_M$ with the model's inputs $\boldsymbol{r}_1 \ldots \boldsymbol{r}_N, s_{z1} \ldots s_{zN}$. If the solution $\psi(\boldsymbol{r}_1 \ldots \boldsymbol{r}_N, s_{z1} \ldots s_{zN}, t)$ to the Schrödinger equation for the underlying quantum counterpart is available, it can produce the solution $\mathcal{F}(q_1 \ldots q_M, t)$ to the given natural-world problem. Evidently, the reduction from finding $\mathcal{F}$ to computing $\psi$ (the quantization procedure) is easy given that it would require a polynomial number of steps (such as quantization of the problem's inputs $q_1 \ldots q_M$ and observations of the initial values of the quantum model's inputs $\boldsymbol{r}_1 \ldots \boldsymbol{r}_N, s_{z1} \ldots s_{zN}$). Consequently, we can say that solving the Schrödinger equation is the **FNP**-complete problem, or, more precisely, it is the problem complete for the class of natural-world problems that is a subset of the **FNP** class.

In this sense, the many-body Schrödinger equation represents the class of natural-world **FNP** problems, since any solution to it can – in combination with the reductions – be used to solve every problem in the class. So, if a fast algorithm could be ever invented for solving the many-body Schrödinger equation in its most general form, then it would have been established that every natural-world problem in **FNP** has a fast algorithm, and, thus, that **FP**=**FNP**.

Therefore, insisting on unrestricted, limitless reductionism (i.e., claiming that every system is basically quantum mechanical) implies the equivalence of the classes **P**=**NP** (as **FP**=**FNP** iff **P**=**NP**[25]), while accepting the limitation of reductionism (in the form of the assumption of limited computational power) would imply the opposite assertion **P**≠**NP**.

In particular, affirming that classical mechanics is simply a quantum mechanics of large systems infers that for each large system the Schrödinger equation has a fast deterministic solution that can be easily transformed in one of classical mechanics. However, this can occur only if the mathematical classes **P** and **NP**-complete are equal. Differently, stating that classical physics does not emerge from quantum physics infers that the Schrödinger equation cannot be efficiently solvable in deterministic terms for every large system describable by classical physics. This necessarily leads to the conclusion that **P**≠**NP**.

As follows, two unresolved problems – the measurement problem in quantum mechanics (that asks whether the quantum description of macroscopic objects can be valid) and the **P** versus **NP** problem in computer science (that asks whether **NP**-complete problems can be solved in polynomial time by a deterministic algorithm) – may possibly be connected. Answer-ing in the negative or the positive to the former might one-to-one determine the answer of "no" or "yes" to the latter, and vice versa.

This means we can phrase the mathematical question of **P** versus **NP** in terms of the quantum measurement problem. And if we accept the physical process criterion for mathematical truth[26], saying that "we should expect a mathematical question to have a definite answer, if and only if we can phrase the question in terms of a physical process we can imagine", this may suggest that the **P** versus **NP** question does have a definite answer.[h]

## 6. Discussion

The assumption of limited computational power is built around the thesis of feasibility of theory, which maintains that in order to be valid a scientific theory is required to be at least feasible. Applying this thesis to the quantum formalism entails that the quantum-mechanical description of a system cannot be valid unless it is based on the feasibly computable solutions to the Schrödinger equation for the system.

Though this thesis seems plain and natural, it is open to as a minimum a few objections.

**1.** The first and possibly the most obvious objection (and yet the most strenuous one) would be that whether a theory is valid or not has nothing to do with the ease of computation. That is, the entire approach undertaken in this paper rests on a basic confusion between the validity of a theory and the practical problem of calculating solutions to its equation of motion.

Before replying to this objection, let us first clarify how the validity of theory is tested. Probably, the most apparent answer would be this: A theory is valid if it can make predictions that can be tested and when tested are found to be true.

---

[h] In other words, this may suggest that the assertion **P=NP** is falsifiable.

The validity of quantum theory for a given physical system rests on the validity of the Schrödinger's equation, which provides a way to calculate all the possible solutions (the wavefunctions) of the system and how they dynamically change in time. This may be express as follows: The quantum description of a system is valid if one can calculate the solutions to the Schrödinger's equation for the given system that can be tested and when tested are found to be true. So, the practical feasibility of calculating the solutions to the Schrödinger's equation is the first and foremost component in the validity of the quantum theory for this system since if the solutions are not feasible, they cannot be testable either.

On the other hand, the theoretical framework of quantum mechanics never calls into question the chances that the Schrödinger's equation has of being solvable for a particular system. This seems to originate in the presupposition that the exact solutions to the Schrödinger equation can always be efficiently calculable (i.e., feasible) for any given *N*-body system. However, this might not be true.

Indeed, if the mathematical classes **P** and **NP**-complete were not equal then solving the Schrödinger equation would be an intractable problem: The equation would be solvable but any exact algorithmic solution to it would run in exponential time (or slower) in the worst case. This means that in the case of a macroscopic system there would not be any real hope of deciding whether this equation is true or false of real experience. Hence, the deterministic quantum-mechanical description of the macroscopic system could not be testable and thus valid.

**2.** When applying the feasibility thesis to the classical formalism, would it entail that the classical-mechanical description of the *N*-body problem of Newtonian gravitation could not be valid too?

Actually, solving the Newton *N*-body problem is computationally easy, because it merely involves integrating the *6N* ordinary differential equations defining the particle motions in Newtonian gravity. This numerical integration is reachable in the number of calculations not greater than $N^2$. Accordingly, one may conclude that the classical-mechanical description of the system of *N* celestial objects interacting with each other gravitationally *can be valid* because it is based on the feasibly computable and so testable solutions to the Newton's equation of motion for the system.

In contrast, solving the quantum *N*-body problem is computationally hard, because in such a system direct (or indirect) interactions between constituent microscopic particles lead to quantum entanglement. As a result, the Schrödinger equation for this system involves a function holding a large amount of information and is therefore difficult, if not impossible, to solve in a reasonable amount of time. For example, a system of only 100 spin-$\frac{1}{2}$ particles requires $2^{100} - 1$ complex numbers to merely describe a general spin state. Consequently, a proof of **P≠NP** would guarantee that the quantum *N*-body problem would be generally intractable and so the deterministic quantum-mechanical description of a truly macroscopic object (consisting of a large number *N* of interacting constituent microscopic particles) could not be valid.

**3.** One need not have solved the Schrödinger equation to show that superpositions of different macroscopic configurations of macro-objects cannot be avoided within a strict quantum mechanical scheme. Therefore, the claim presented in the paper that the difficulty of obtaining a solution to the Schrödinger equation for macro-objects could resolve the measurement problem is unsustainable.

Presumably this objection results from the following logic: Since at the microscopic level the Schrödinger equation has the testable property of linearity (that has been experimen-tally verified extensively through the observation of interference effects, for instance) and since during transition from the microscopic level to the macroscopic one the form of this equation is in no way changed, then its property of linearity cannot have changed either.

However, such an assumption ignores the inherent computational hardness of the Schrödinger equation. If **NP**-complete problems had no deterministic polynomial-time solutions (i.e., **P≠NP**), then at the macroscopic level the Schrödinger equation – even though its form would not be changed – would become unsolvable in deterministic terms for all practical purposes. This means that exact macroscopic solutions to the Schrödinger equation as well as their linear combinations would not be feasible and thus testable.

**4.** The difficulty of obtaining a solution to the Schrödinger equation depends on the choice of basis, in which the wave function is expressed (in the paper the choice is positions and spin). But by choosing a different basis, i.e., applying a unitary transformation, the equation might become much simpler. (This corresponds to the transition from the Schrödinger to the Heisenberg representation.) It is not clear how the dependence of computational difficulty on representation would bear on the claim about the validity of the theory, as implied by the feasibility thesis, assuming that the validity of a theory is independent of its representation.

It is true that by certain manipulations, such as coordinate transformations, the choice of a different basis, or a basis change with the respect to time-dependency (e.g., the transition from the Schrödinger picture to the Heisenberg one), it is possible to make the *N*-body Schrödinger equation easier to handle (or even to solve it) in some special cases and for some limited *N*.

The trouble is, however, that even though such manipulations that bring down the *N*-body Schrödinger equation has been rightly guessed in some cases, no particular rule is followed how to have the same lucky guess in all other cases. Since this rule is unknown, to search through possible manipulations one has to incline to the use of brute-force methods, which get infeasible very quickly as *N* grows. This is why the *N*-body Schrödinger equation is computationally difficult, and this difficultness (**FNP**-complete) is independent from the particular representation or basis.

**5.** How would the computational difficulty depend on the interaction between the particles? If, for example, the solution to the Schrödinger equation becomes intractable regardless of the interaction term, i.e., even if there is no interaction, then quantum theory would seem to break down in the macroscopic limit regardless of whether a measurement is being made. That would not be a solution of the measurement problem per se.

Perhaps, the most important question in the measurement problem is how to describe the interaction between the microscopic quantum object and the (macroscopic) measuring apparatus. Specifically, should the measuring apparatus be described as a quantum system and hence all the measuring process as an interaction between two quantum systems?

The answer to this question (and consequently the resolution of the measurement problem) depends on the answer to the **P** versus **NP** question.

If, for instance, **P**≠**NP**, it entails that the measuring apparatus is infeasible to describe by exact solutions to the apparatus's Schrödinger equation. Consequently, if the microscopic quantum object (initially isolated and described by the feasibly computable deterministic solutions to the micro-object's Schrödinger equation) were brought into interaction with the measuring apparatus, then after the measurement interaction the only solutions to the Schrödinger equation for the micro-object feasible to compute would be probabilistic solutions.

On the other hand, if **P**=**NP**, it results in that the quantum formalism should be taken to apply to all physical systems, including the measuring apparatus. In that case, if such an apparatus describable by feasibly computable exact solutions to the apparatus's Schrödinger equa-tion interacted with the microscopic quantum object, the final state of these two quantum systems (the micro-object + the apparatus) could be a superposition.

**6.** The claim that solutions of the many-body Schrödinger equation are in general unfeasible, i.e., not reachable in polynomial time, is only an assumption of the paper, but would need substantiation.

Though in the paper the **FNP**-completeness of the general deterministic solution to the many-body Schrödinger equation is presented as rather a theoretical conclusion drawn from abstract reasoning, it can be easily substantiated by the known facts about the computational hardness of this equation in many particular cases.

For example, the efficient (i.e., polynomial-time) general solution of the Schrödinger equation for a system with large number of strongly interacting fermions would imply the full and generic solution of the numerical sign problem – the well-known **NP**-hard problem[27] – in polynomial time.

Furthermore, to provide a deterministic quantum-mechanical description of a truly macroscopic object would involve a precise determination of its molecular binding energy, which in turn would require the efficient solution to the so-called *N*-representability problem. However, as it has been shown recently[28], both fermionic and bosonic versions of this problem are **QMA**-hard (referring to quantum Merlin-Arthur problems, which are considered difficult even for quantum computers).

**7.** Carroll[29] raises this objection: "No one denies that in practice we can never describe human beings as collections of electrons, protons, and neutrons obeying the Schrödinger equation. But many of us think that this is clearly an issue of practice vs. principle; the ability of our finite minds to collect the relevant data and solve the relevant equations should not be taken as evidence that the universe is not fully capable of doing so."

According to the survey[20], none of the models of computation – known or potential one – that are based on physical processes can solve **NP**-complete problems in polynomial time. So, the computational hardness of the Schrödinger equation looks more like the naturally imposed limit to reductionism than our own inability as human beings to solve difficult problems.

**7. Conclusion**

In nearly all of the papers on the foundations of quantum mechanics, when discussing the measurement problem and its possible solutions, the authors never doubt ability of the Schrödinger equation to be exactly solved for every physical entity (regardless of spatial size, physical property, and the like), including "all environmental particles within the causal horizon for the apparatus"[30], "observers with their perceptive and cognitive apparatuses"[11], and even the entire universe. Such a belief in limitless reductionism, i.e., in the across the board efficient solvability of the Schrödinger equation, is exactly what makes possible for one to declare a truly macroscopic object to be a quantum one (for the further use in the theory) and thus to allow the occurrence of superpositions of macroscopically different states of the macro-object.

Apparently, the reasoning behind this belief is the following: It does not matter that at this point in time the equation of motion of a theory is much too complicated to be exactly soluble for a given system because even if the required solution cannot be ultimately expressed as a closed-form or analytic expression, a sufficiently powerful numerical method will always be able to compute it with an arbitrary precision and in reasonable time.

Seeing that this line of reasoning works well in the case of the equation of motion of classical or relativistic mechanics, one might expect that the same argument would also be justifiable in the case of the Schrödinger equation.

However, as the recent developments in the computational complexity theory imply, this is an unlikely hypothesis. Since the standard complexity theorists' opinion is that there are no deterministic polynomial-time solutions to **NP**-hard problems (though proving this seems as one of the deepest problems in all of mathematics), the many-body Schrödinger equation is an intractable problem, for which there are not any exact generic polynomial-time solutions.

This may suggest a likely resolution to the measurement, or macro-objectification, problem: The purely quantum description is not applicable to a truly macroscopic object because the Schrödinger equation for this object is unsolvable except in terms of probability for all practical purposes.